\documentclass[12pt]{iopart}
\usepackage{iopams} 
\usepackage{graphicx} 
\begin{document}

\title{Effect of chemical substitution in Rh$_{17}$S$_{15}$}

\author{Naren H.R., Arumugam Thamizhavel and Ramakrishan S.}

\address{Tata Institute of Fundamental Research, Mumbai, India}
\ead{nareni@tifr.res.in}

\begin{abstract}
Rh$_{17}$S$_{15}$ has recently been shown to be a strongly correlated superconductor with a transition temperature of 5.4 K. In order to understand the nature of strong correlations we study the effect of substitution of some of the Rh and S atoms by other elements such as Fe, Pd, Ir and Ni on the Rh side and Se on the S side in this work. We find that while substitution of Ir and Se lower the transition temperature considerably, that of Fe, Pd and Ni destroy superconductivity down to 1.5 K. The resistivity data in these doped samples show a minimum which is presumably disorder induced. A reduction of T$_c$ is always accompanied by a reduction of electron correlations as deduced from heat capacity and magnetization data. Interestingly, the Fe doped sample shows evidence for a spin glass formation at low temperatures.
\end{abstract}

\maketitle

\section{Introduction}
\begin{figure*}
	\centering
		\includegraphics[width=12cm]{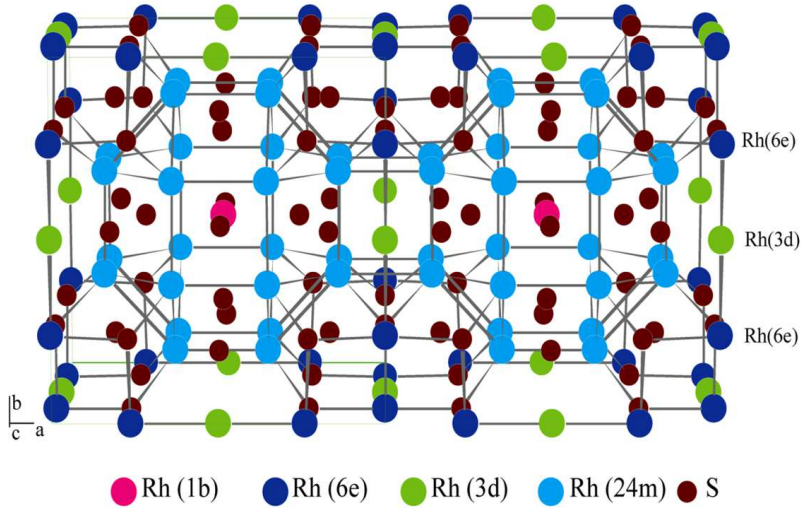}
	\caption{{[colour online] The crystal structure of Rh$_{17}$S$_{15}$ showing two unit cells.}}
	\label{struct}
\end{figure*}

Rh$_{17}$S$_{15}$ was remarkable for its strong electron correlations as evidenced by an enhanced Sommerfeld coefficient, susceptibility and upper critical field values as shown by us(Ref.~\cite{r1}). It is a rare example of a strongly correlated superconductor which is 4d electron based (Sr$_2$RuO$_4$ being the other known system). Its cubic crystal structure (space group Pm3m), shown in fig. \ref{struct}, shows Rh occupying four and S occupying three different crystallographic sites respectively. Also the structure has large vacancies between two unit cells (at the 4c sites) which could accommodate dopant elements and thereby change the physical properties of the system. This was a reason for our initial interest in this compound. Another compelling reason to study the effect of dopants on this system is our conjecture (Ref.~\cite{r1}) that the origin of the strong electron correlations in Rh$_{17}$S$_{15}$ could be due to a narrow Rh d band at the Fermi level which arises because of closely lying Rh atoms in the crystal structure. A natural way to verify this would be to study the effect of pressure, either chemical or physical, on this system. 

Recently, Settai et al (Ref.~\cite{r19}) showed that application of pressure upto 2.7 GPa increases the density of states at the Fermi level (as deduced from the slope of the upper critical field at T$_c$) but reduces the T$_c$ and the upper critical field values. 

We have tried a large number of elemental substitutions on this compound and find that it is difficult to predict which element will get into the system while maintaining the crystal structure of Rh$_{17}$S$_{15}$ and not form newer phases. Also, it is not possible to predict before hand whether the Rh or S atoms are going to get substituted or whether the dopant atoms will get into the interstices. Our approach has been to grow the crystals with the charge taken in the form Rh$_{16}$X$_1$S$_{15}$ and Rh$_{17}$S$_{14}$Y$_1$ where X and Y are the dopant atoms. Then perform a powder XRD measurement and determine the crystal structure and lattice parameters by a Reitveld analysis. Once we find that the sample has formed in the desired crystal structure without any prominent impurity peaks we cross-verify with an Electron Probe Microanalysis (EPMA) to study the elemental composition  at different regions of the sample. The samples are selected only once both the above techniques reveal that it is an essentially single phase compound. Using this approach, we report here a study of substitutions of the type Rh$_{16}$X$_1$S$_{15}$ and Rh$_{17}$S$_{14}$Y$_1$ where X = Fe, Pd, Ir and Ni and Y = Se. Inevitably, we have tried many more dopants and some (Mo, Nb, Ru, Ti, V, Ca, Sc and Co) of them have been found to not form in the desired crystal structure while the rest (Ag, Mg and Te) do not enter the crystal structure of Rh$_{17}$S$_{15}$.  

We have also reported (Ref.~\cite{r2}) the effect of Iridium substitution for Rh (Ir$_x$Rh$_{17-x}$S$_{15}$), for x = 1 and 2, where we found that an increased size of the unit cell reduced the electron correlations.

\section{Experimental details}
\begin{figure*}
	\centering
		\includegraphics[width=12cm]{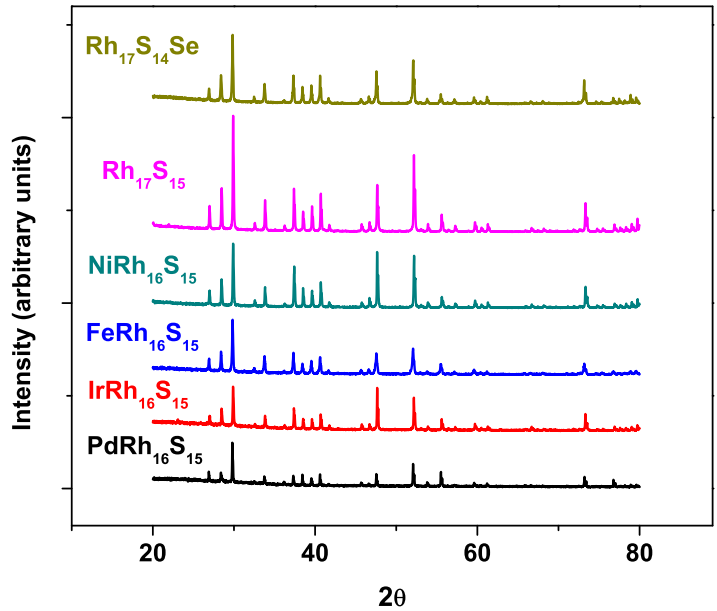}
	\caption{{[colour online] The powder XRD plots of the doped samples along with the undoped one}}
	\label{xrd}
\end{figure*}

High purity powders (99.99\% and better) of the relevant elements were mixed and pelletised in stoichiometric ratios as determined by Rh$_{16}$X$_1$S$_{15}$ and Rh$_{17}$S$_{14}$Y$_1$. These pellets were put in conical base alumina crucibles, evacuated to a vacuum of around 10$^{-6}$ mBar and sealed in a quartz tube. This charge was heated slowly (8 $^{\circ}$C/hr) to 1150 $^{\circ}$C and then annealed at 1080 $^{\circ}$C for two days. Then cooled to 600 $^{\circ}$C at 8 $^{\circ}$C/hr and then rapidly cooled down. We obtained polycrystals of doped samples in this method. Powder XRD was performed on these samples and the pattern was compared with that of Rh$_{17}$S$_{15}$ (fig. \ref{xrd}). The ones which do not have any impurity peaks were selected and refined using a Reitveld analysis. After this analysis, we could decide whether the unit cell volume had increased or decreased which in turn meant negative or positive chemical pressure respectively. All the doped samples with the exception of the Ni doped one show an increase in the lattice constant and hence imply a negative chemical pressure. We have tabulated the lattice constants in the table (fig. \ref{dopedtable}). 

We have also performed Electron probe micro analysis (EPMA) based elemental analysis of these selected samples over areas as large as 2 mm * 2 mm. This phase sensitive analysis shows a fairly uniform distribution of dopant atoms. While from the initial stoichiometry one expects around 3 atomic percent of dopant in every region, we find a distribution of 1.5 to 4 atomic percent in all the samples except in the Ni doped sample. The Ni doped sample shows a distribution of 0.3 to 2 atomic percent and some small regions of excess Ni aggregation. Even in the other samples, there are very small regions which have a stoichiometry of a different phase. From the absence of impurity peaks in the XRD patterns and the EPMA analysis we conclude that the crystal structure of Rh$_{17}$S$_{15}$ has been essentially maintained in the doped samples and that there are no major stray phases formed. Also, since Rh occurs in 4 and S in 3 different crystallographic sites, it is not possible from this study to decide which Rh or S site is being substituted.

We performed magnetization measurements in a squid magnetometer (Quantum Design, USA), resistivity measurements in home-made resistivity inserts and specific heat capacity measurements in a Quantum Design PPMS.

\section{Results and Discussion} 
We describe the experimental results on all the doped samples except the Fe doped one in this section. The Fe doped sample results are described as a separate section.

\subsection{Resistivity} 
\begin{figure*}
	\centering
		\includegraphics[width=12cm]{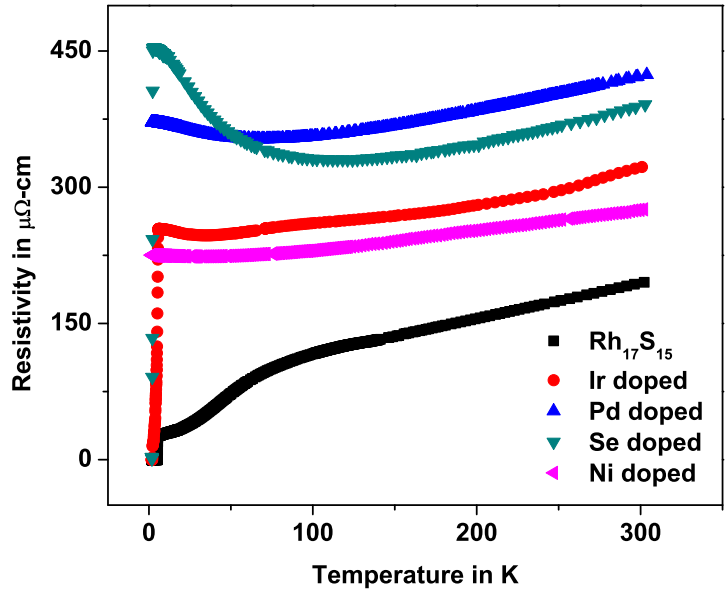}
	\caption{{[colour online] The zero field resistivity curves of the doped samples as contrasted with undoped Rh$_{17}$S$_{15}$. Notice the minima in the doped curves.}}
	\label{rhos}
\end{figure*}

The resistivity data of the doped samples are very dissimilar to that of undoped Rh$_{17}$S$_{15}$. In fig. \ref{rhos} we have displayed the data of undoped Rh$_{17}$S$_{15}$ as compared with the doped samples all measured in the absence of a magnetic field. Other than Ir and Se doped samples none of the rest show superconductivity down to 1.5 K. While in the Ir doped sample the transition is rather broad (starts at 5 K and finally becomes superconducting below 2.2 K) the transition is sharp in the Se doped sample and it is superconducting below 2.2 K. The most distinct feature in the doped curves is the appearance of a minimum (at different temperatures for different samples). The temperature of this minimum (T$_{min}$) is independent of magnetic field. From the table (fig. \ref{dopedtable}) we can see that T$_{min}$ increases as the lattice constant increases. 

Such a minimum in resistivity has been observed in a wide variety of systems starting from amorphous and glassy systems (Ref.~\cite{r3,r4,r5,r6,r7}) to poly crystalline and single crystalline metals (Ref.~\cite{r8,r9,r10,r11}) to oxides (Ref.~\cite{r12,r13}) to nanostructured materials (Ref.~\cite{r14}). Early explanations for this phenomenon was by invoking the Kondo effect. However, this approach was rejected largely on the grounds of the insensitivity of this minimum to a magnetic field (Ref.~\cite{r15}). An alternative explanation was based on a two-level system (TLS) approach where ions free to tunnel between two vibrational levels opened up newer scattering channels and hence an increase in resistivity at low temperatures (Ref.~\cite{r4}). This mechanism is expected to be prominent in disordered materials like glasses and is supposed to disappear on crystallization. 
Another approach to this problem, due to Altshuler and Aronov (Ref.~\cite{r16}), suggests that the semi-classical approach to transport breaks down in the presence of strong disorder. Matheissen's rule stating that effect of multiple scatterers can be summed up individually taken one at a time needs to be reconsidered. That is the interference losses due to waves from different scatterers cannot be neglected. Such quantum corrections to the resistivity gives a -T$^{0.5}$ term in the resistivity at low temperatures. Generally, the resistivity can be represented as
$$\rho = \rho_{el} + \rho_{in}  \eqno(1)$$
where $\rho_{el}$ is the elastic scattering contribution due to electron-impurity interaction and electron-electron Coulomb interaction and $\rho_{in}$ is the inelastic scattering contribution due to electron-phonon interaction. $\rho_{in}$ is expected to increase as a power law bT$^c$ where `b' and `c' are independent of magnetic field. In good conductors $\rho_{el}$ is expected to be independent of temperature and amounts to a residual resistivity term $\rho_0$. However, with strong disorder there is a correction to this term which has a -T$^{0.5}$ dependence. So the effective low temperature behaviour follows 
$$\rho(T) = \rho_0 - aT^{0.5} + bT^c   \eqno(2)$$
where the coefficient `a' is expected to be strongly dependent on the amount of disorder and hence on $\rho_0$ (Ref.~\cite{r12}). Such an dependence will give a minimum of resistivity at a characteristic temperature T$_{min}$ which represents the scale at which the resistivity changes from disorder dominated to phonon dominated. The coefficient `a' is expected to be proportional to $\rho_0^2$.
The Resistivity can be fitted fairly well with the above equation. We have fitted the data only  about the minimum (0.5 T$_{min}$ $\textless T $\textless 1.5 T$_{min}$) since the low temperature data could be affected by superconductivity in some residual Rh$_{17}$S$_{15}$ phase. The obtained coefficients are tabulated in the table (fig. \ref{dopedtable}). Clearly, T$_{min}$ increases as the residual resistivity ($\rho_0$) increases. The coefficient `a' representing the disorder induced correction to $\rho_0$ also increases with $\rho_0$. By extrapolating a straight line fit to T$_{min}$ and `a' versus $\rho_0^2$ it is seen that the value of $\rho_0$ for which both T$_{min}$ and `a' vanish is around 260 $\mu\Omega$-cm which is a measure of the amount of disorder necessary for quantum corrections to resistivity to start becoming significant. Another observable trend is the decrease of the power `c' of the increasing term in equation 2 with increase in disorder.
 
\begin{figure*}
	\centering
		\includegraphics[width=12cm]{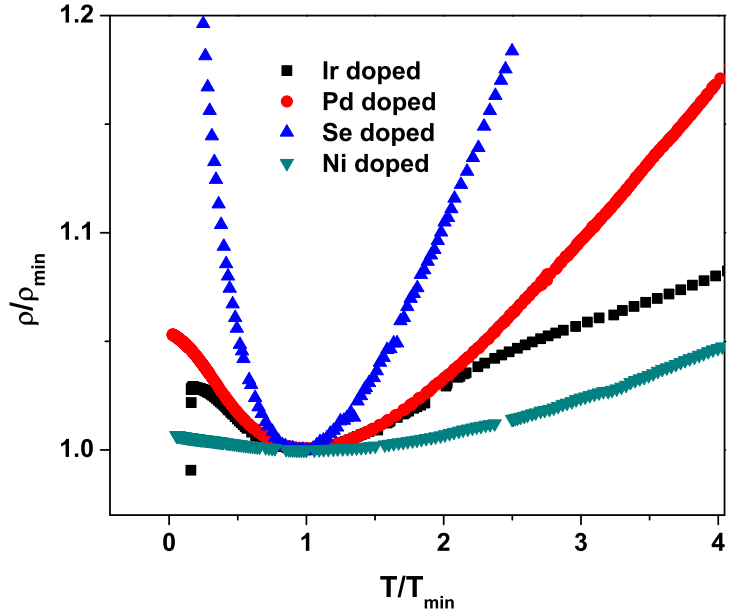}
	\caption{{[colour online] The scaled zero field resistivity curves of the doped samples.}}
	\label{scaledrhos}
\end{figure*}

A study of varying disorder in NbS$_2$ (Ref.~\cite{r8}) shows a scaling of the resistivity data when T is scaled by T$_{min}$ and $\rho$ by $\rho_{min}$. In fig. \ref{scaledrhos} we have plotted the scaled resistivity data. We do not see a scaling as seen in ref. ~\cite{r8}. The reason could be that this scaling is possible when we are dealing with different quantities of the same dopant rather than different dopants like in our case. However, we do see a reasonable collapse of data of Ir and Pd doped data in a small range about the minimum. 
 
Another prediction of this theory (Ref.~\cite{r16}) is that the disorder causes a depression of density of states (DOS) at the Fermi level. We will comment on this in the concluding section.

\subsection{Susceptibility}
\begin{figure*}
	\centering
		\includegraphics[width=12cm]{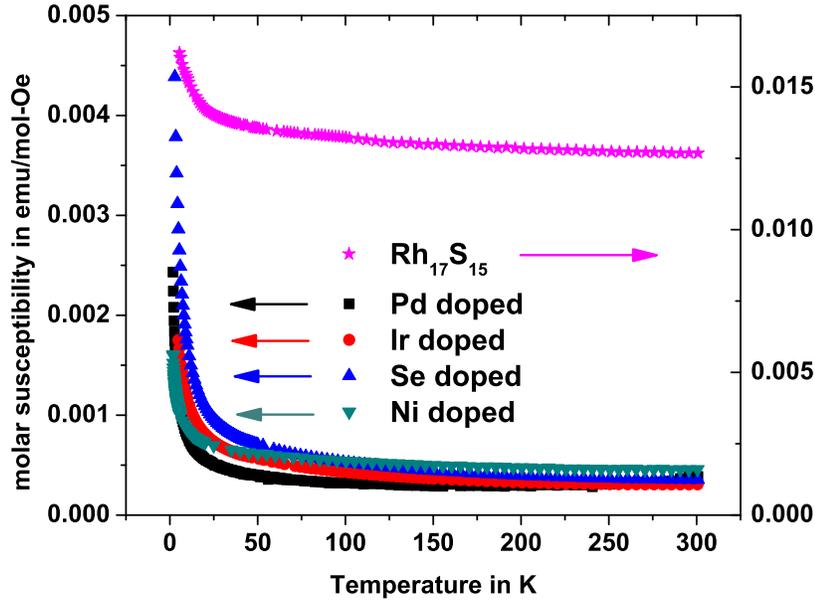}
	\caption{{[colour online] The susceptibility curves (in a field of 1000 Oe) of the doped samples plotted along with that of undoped Rh$_{17}$S$_{15}$. Notice the different Y axes for the doped and undoped curves.}}
	\label{dcsus}
\end{figure*}

The susceptibility curves of the doped samples (in a field of 1000 Oe) are plotted along with that of undoped Rh$_{17}$S$_{15}$ in fig. \ref{dcsus}. Notice the different Y axes for the doped and undoped curves. The curves quantitatively look the same wherein we see a small rise at low temperatures. We believe that Fermi level motion in these narrow band systems cause such a  temperature dependent Pauli susceptibility as suggested for V$_3$Si (Ref.~\cite{r20}). However, the values of susceptibility are much lower in the doped samples. The room temperature susceptibility which is a direct measure of the DOS at Fermi level is around two orders of magnitude smaller in the doped samples when compared with the undoped one (see table (fig. \ref{dopedtable})). 
 
\subsection{Heat Capacity}
The low temperature heat capacity (C$_p$) data of all the doped samples was fitted to a                                               
$$ C_p~=~ \gamma~T+~\beta~T^{3}~+~\delta~T^5 \eqno(3) $$ 
where $\gamma$ is due to the electronic contribution, $\beta$ is due to the lattice contribution and $\delta$ is the contribution due to anharmonicity. We recall here that the Sommerfeld coefficient ($\gamma$) of Rh$_{17}$S$_{15}$ was 108.4 mJ/mol-K$^2$. In comparison the Sommerfeld coefficents of the doped samples (see table (fig. \ref{dopedtable})) are reduced considerably which again indicates a reduction in the DOS at Fermi level. The Debye temperatures ($\theta_D$), however, are only slightly reduced except for the Ir doped sample where the reduction is considerable. This reduction implies a softening of the lattice as compared with the undoped system.

\subsection{Upper Critical Field}
In the superconducting Ir and Se doped samples the upper critical field is considerably reduced. The upper critical field (H$_{c2}$(0)) in Rh$_{17}$S$_{15}$ is around 20 T. In Ir doped sample it reduces to around 12 T and is further reduced to around 6 T in Se doped sample. Since the upper critical field increases with the DOS at Fermi level  (H$_{c2}$(0) $\sim$ DOS$^2$), this is another indicator of the reduction in DOS. 

\section{F\lowercase{e doping}}
Iron doping causes dramatic changes in the properties to the extent of introducing some magnetic ordering as well.

\begin{figure*}
	\centering
		\includegraphics[width=12cm]{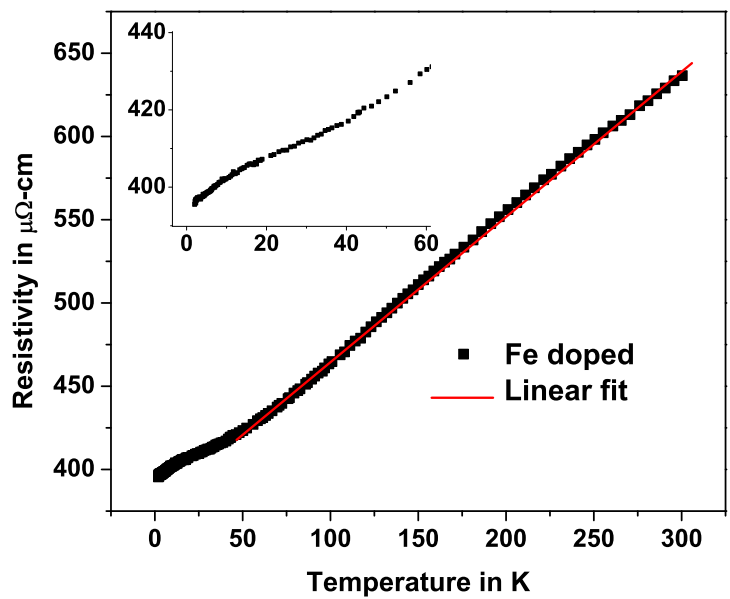}
	\caption{{[colour online] The zero field resistivity curve of Fe doped sample. The continuous line is a linear fit above 50 K }}
	\label{ferho}
\end{figure*}

The zero field resistivity curve (fig. \ref{ferho}) shows an extended linear region above 50 K and the fit is shown by a continuous red line. Large ranges of linear resistivity has been observed in Rh-Fe systems (Ref.~\cite{r17,r18}) where Fe is present in very small amounts (\textless~ 1\%). It is also speculated that larger amounts of Fe may lead to magnetic ordering. In our system we have a distribution of 1.5 to 4 atomic percent of Fe in place of Rh in Rh$_{17}$S$_{15}$ and we do find hints of magnetic ordering. In the inset of fig. \ref{ferho} we show the low temperature region (\textless~ 50 K) which shows a stronger decrease in resistivity below 15 K. This could be due to setting in of some kind of magnetic order for which further support comes from the magnetization data. There is no significant effect of application of magnetic field of upto 1 T on the resistivity.

\begin{figure*}
	\centering
		\includegraphics[width=12cm]{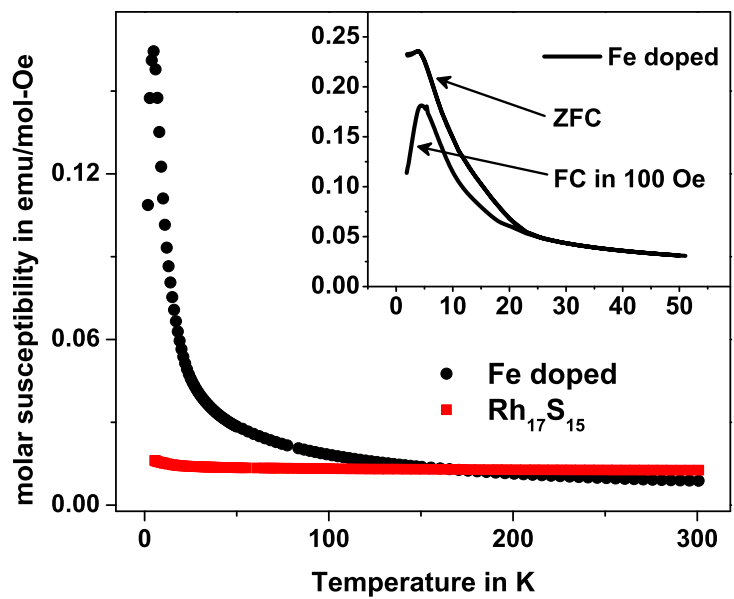}
	\caption{{[colour online] The dc susceptibility curve of Fe doped sample as contrasted with the undoped Rh$_{17}$S$_{15}$. Notice the downturn in the Fe doped curve at low temperatures. Inset contains the Zero Field Cooled (ZFC) and Field Cooled (FC) curves}}
	\label{fezfc}
\end{figure*}

\begin{figure*}
	\centering
		\includegraphics[width=12cm]{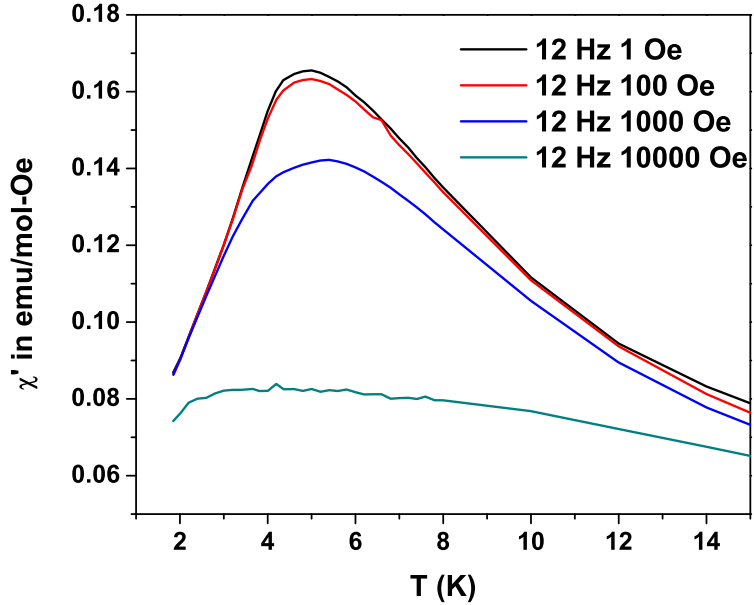}
	\caption{{[colour online] AC susceptibility ($\chi$') is plotted as function of temperature at different fields around the downturn region. The downturn clearly broadens with increase in field}}
	\label{fielddep}
\end{figure*}

\begin{figure*}
	\centering
		\includegraphics[width=12cm]{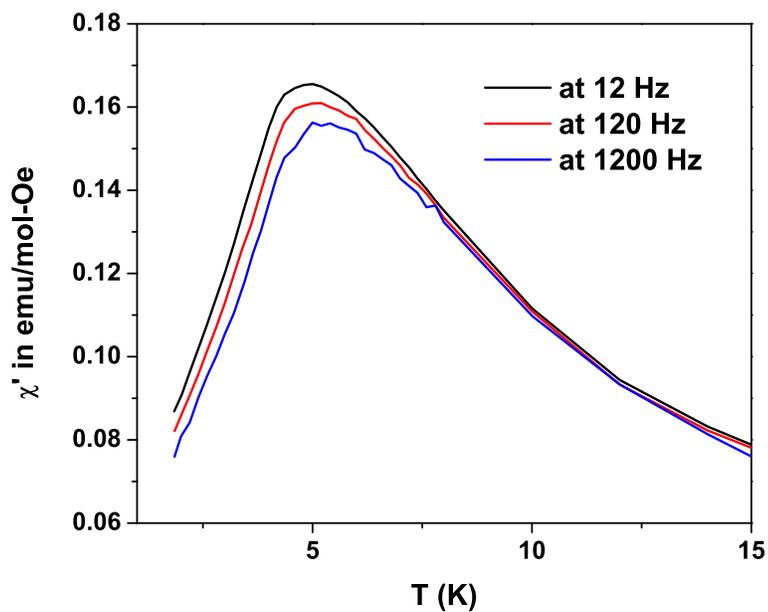}
	\caption{{[colour online] AC susceptibility ($\chi$') is plotted as function of temperature at different frequencies around the downturn region. The downturn clearly shows a frequency dependence}}
	\label{freqdep}
\end{figure*}

The dc susceptibility curve is shown in fig. \ref{fezfc} as compared to that of Rh$_{17}$S$_{15}$. While the room temperature susceptibility of the Fe-doped sample is lower than that of Rh$_{17}$S$_{15}$, at lower temperatures the susceptibility of  Fe-doped system becomes much larger and the largest value is an order of magnitude larger than the undoped system. The inset of fig. \ref{fezfc} shows the zero field cooled and field cooled curves showing a departure from each other below 20 K suggesting the setting up of magnetic order. Also, a marked downturn is seen in the data at low temperatures which could either be attributed to setting up of antiferromagnetic order or the formation of a spin glass. To check for the latter we performed ac susceptibility measurements at different fields and at different frequencies around the downturn region as shown in fig. \ref{fielddep} and fig. \ref{freqdep} respectively. The ac field amplitude was 1 Oe and the sample was cooled in zero field. Clearly the $\chi$'(T) data show a broadening of the downturn with increase in field (fig. \ref{fielddep}) and a frequency dependence of the onset of the downturn. Also notice that the height of the peaks reduce with increase in field and frequency. These features are the defining features of the formation of a spin glass (Ref.~\cite{r21}). 

\begin{figure*}
	\centering
		\includegraphics[width=12cm]{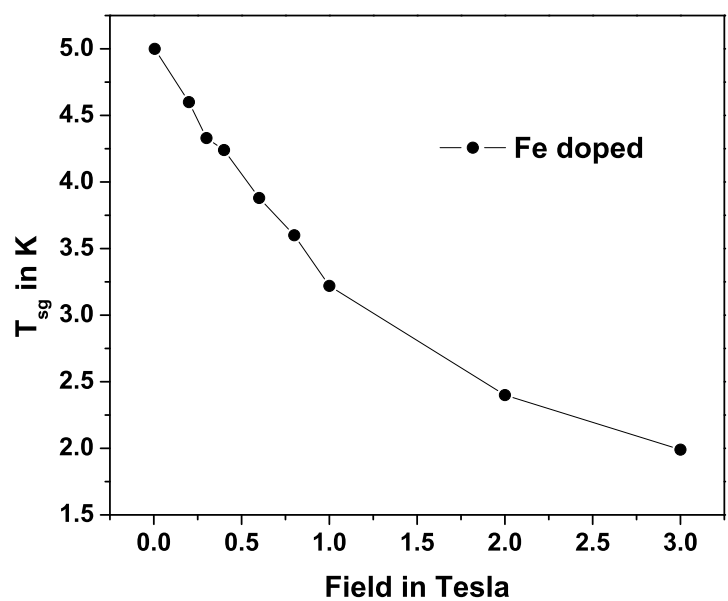}
	\caption{{[colour online] T$_{sg}$, as deduced from the downturn in the dc susceptibility data, plotted against magnetic field.}}
	\label{tsg}
\end{figure*}

Hence we can conclude that the Iron spins freeze below a certain temperature (T$_{sg}$), the temperature below which the susceptibility starts to decrease. In fig. \ref{tsg} we have plotted T$_{sg}$, as a function of field and clearly T$_{sg}$ decreases with increase in field. 

\begin{figure*}
	\centering
		\includegraphics[width=12cm]{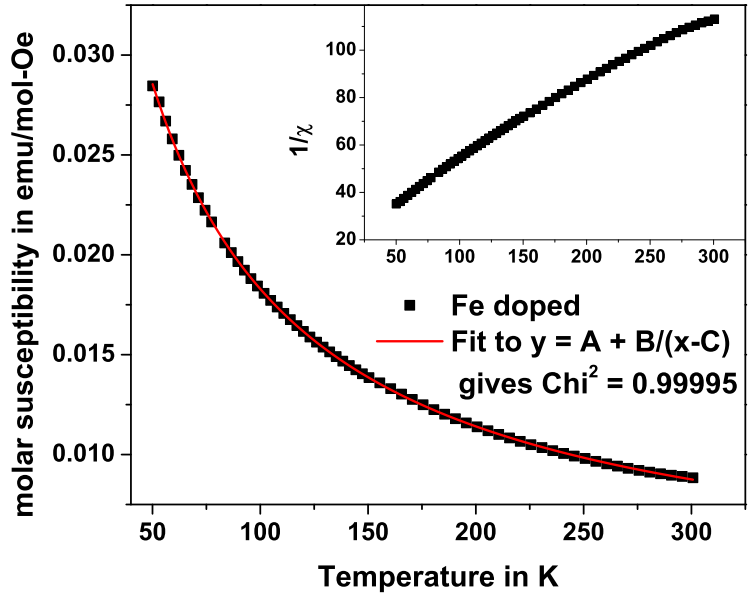}
	\caption{{[colour online] Curie-Weiss fit of the susceptibility data above 50 K. The fit values are discussed in the text. The inset shows $\chi^{-1}$ versus T which is not perfectly linear}}
	\label{curie}
\end{figure*}

In order to estimate the value of moment in the paramagnetic phase, we fit the susceptibility above 50 K to a Curie-Weiss equation as shown in fig. \ref{curie}. From this fit we get an 'A' of 0.0028 emu/mol-Oe, 'B' of 1.95 emu*K/mol-Oe and 'C' of -25.5 K. From the value of 'B', we can estimate a moment of 1.25 Bohr magneton per formula unit. We recall here that from the substituted stoichiometry we expect one Iron atom per formula unit on an average. So we can conclude that an average Iron spin in the sample has a moment of 1.25 Bohr magneton. The inset of fig. \ref{curie} shows $\chi^{-1}$ plotted against T. Clearly the behaviour is not perfectly linear. This could be due to the fact that the value of 'A' from the fit is quite large.

\begin{figure*}
	\centering
		\includegraphics[width=12cm]{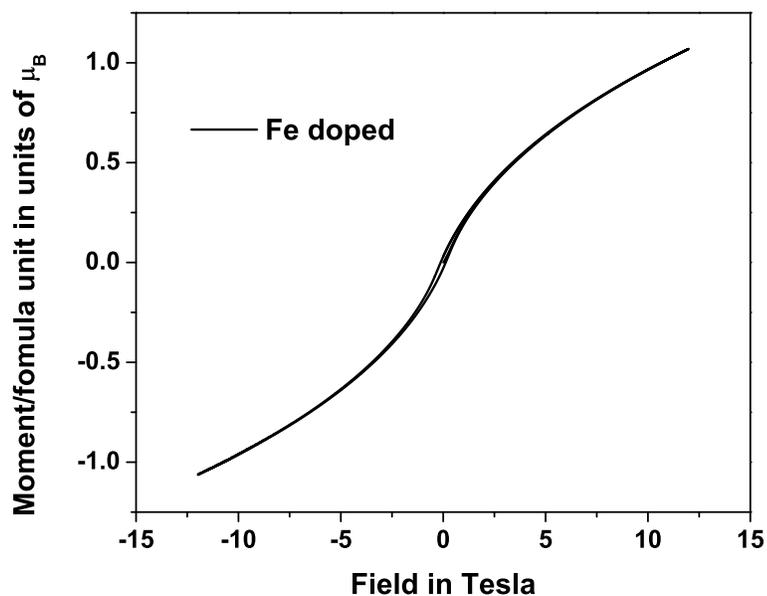}
	\caption{{[colour online] The hysteresis loop of Fe doped sample (at 1.7 K) showing a small hysteresis below 5 T and linear behaviour beyond.}}
	\label{mhloop}
\end{figure*}

In fig. \ref{mhloop} we have the five quadrant hysteresis loop (at 1.7 K) where we have plotted the molar moment per formula unit of Rh$_{17}$S$_{15}$ in units of Bohr magneton ($\mu$B). We see a very small hysteresis at fields lesser than 5 T and beyond that a linear response. The hysteresis is due to the formation of a spin glass. The linear part seen beyond 5 T is because the sample becomes paramagnetic in such large fields at 1.7 K. 

\begin{figure*}
	\centering
		\includegraphics[width=12cm]{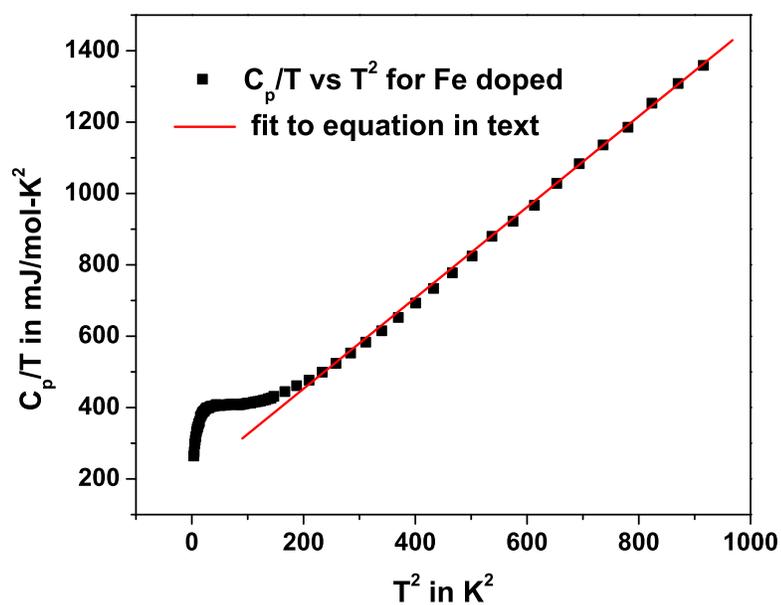}
	\caption{{[colour online] C$_p$/T, measured in zero field, plotted against T$^2$. The continuous line is a fit to the equation described in the text. Notice the change in behaviour below 14 K.}}
	\label{fecp}
\end{figure*}

Fig. \ref{fecp} shows the low temperature (zero field) specific heat capacity data of Fe doped Rh$_{17}$S$_{15}$ where C$_p$/T is plotted against T$^2$. In metallic systems following Debye theory we expect a linear behaviour. In this system we observe a linear behaviour only at temperatures above 14 K. There is a rapid reduction of entropy below 6 K. This we attribute to setting up of spin glass like order in the system. A fit to equation 3 in the linear region gives us an enhanced Sommerfeld coefficient of 223 mJ/mol-K$^2$ which is more than double the value in the undoped system. This could be due to enhanced electron correlations due to setting in of magnetic order in the system. The Debye temperature is reduced to 378 K in the Fe doped sample. 

\section{Conclusions}

\begin{figure*}
	\centering
		\includegraphics[trim=0cm 17.5cm 0cm 0cm, clip=true, width=16cm]{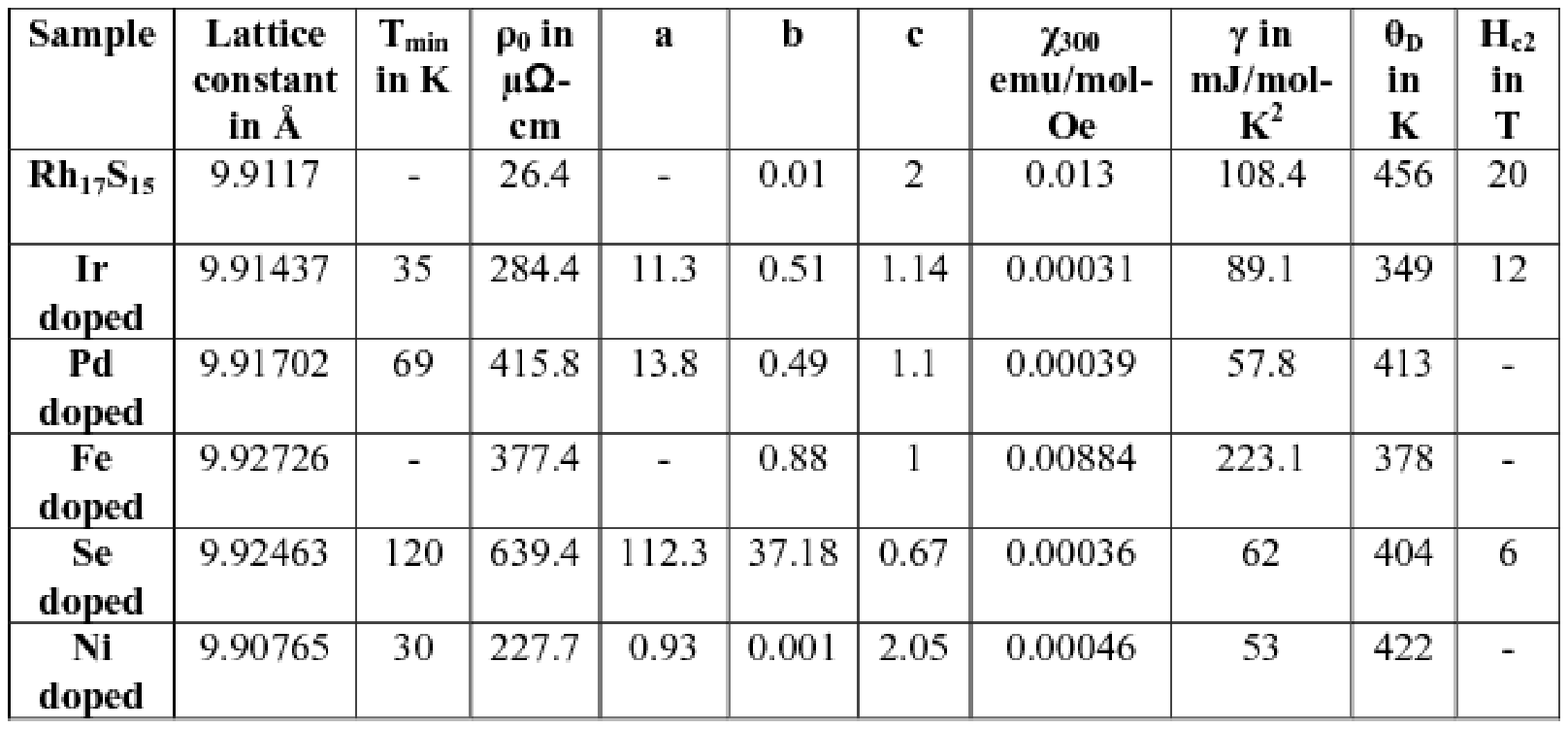}
	\caption{{[colour online] table containing the calculated parameters of the doped samples and the undoped Rh$_{17}$S$_{15}$. The columns titled a, b and c are the coefficients to the resistivity minimum fits as described in the text.}}
	\label{dopedtable}
\end{figure*}

The table (fig. \ref{dopedtable}) provides for a comparison of various parameters discussed in the earlier sections. The resistivity data in the doped samples can be understood in terms of effects of disorder. Disorder causes a minimum in the resistivity and we find that by fitting our data to the suggested equation we obtain a value of around 260 $\mu\Omega$-cm as the residual resistivity necessary for quantum corrections to resistivity to become significant. The reduced DOS at Fermi level which is reflected in the reduced susceptibilities, Sommerfeld coefficients and upper critical field values could be due to two reasons. Firstly, it could be due to a suppression of electron correlations in the doped samples due to increases Rh-Rh distances due to expansion of the unit cell as seen by the increase in the lattice constant. However, the Ni doped sample, which again shows a reduction of DOS at Fermi level, shows a mild contraction of the unit cell. This is something we do not expect particularly in the light of the work of Settai et al (Ref.~\cite{r19}) where an application of pressure on Rh$_{17}$S$_{15}$ caused an increase in DOS at Fermi level. A more concentrated and systematic effort towards chemical substitution is planned which should offer more insights into the effect of positive chemical pressure. The increase of disorder is also expected to cause a reduction of DOS at Fermi level. In fact, it is supposed to cause a minimum at the Fermi level. This could be verified by DOS measurements by photoemission spectroscopy and this work is in progress. A minimum, if observed would provide more support to the notion that it is the disorder that plays a dominant role in the reduction of DOS at Fermi level. 

In the case of the Iron doped system there is evidence for formation of a spin glass at low temperatures from the susceptibility data and its possible signatures in the resistivity and heat capacity data. Since the distances between Iron atoms are not uniform it could cause a frustration in the ensuing RKKY interaction between the Iron moments. Since it is a doped system that we are dealing with there is a lot of disorder in the crystal as well. Hence the two canonical requirements for the formation of a spin glass (Ref.~\cite{r21})- frustration and disorder - are both present in this system.

The fact that superconductivity survives in the Se and Ir doped samples and is absent in the others bears no correlation either with the amount of disorder or with the increase in lattice constant. While magnetism in the Fe doped sample could kill the superconductivity there is no apparent magnetism observable in the Ni and Pd doped samples. This is something we do not understand at present. Also, the fact that superconductivity is hampered by non-magnetic impurities like Se, Ir and Pd is interesting since, for conventional BCS superconductors non-magnetic impurities should not affect the superconductivity. However, it has been widely observed in the Heavy Fermion superconductors that superconductivity is strongly affected by non-magnetic impurities (Ref.~\cite{r22}). This is attributed to the presence of non s-wave gap symmetry in these systems. Efforts to determine the gap and its symmetry in Rh$_{17}$S$_{15}$ are on presently.

\ack
We would like to acknowledge the help of Ms. Sangeetha NS and Dr. Ravi Prakash Singh in some of the measurements.


\section*{References}
\begin{thebibliography}{10}
\bibitem{r1} H.R. Naren, A. Thamizhavel, A. K. Nigam, and S. Ramakrishnan 2008 {\it Phys. Rev. Lett.} {\bf  100} 026404
\bibitem{r2} H.R. Naren, A. Thamizhavel, A. K. Nigam, and S. Ramakrishnan 2010 {\it Physica C} {\bf  470} 682
\bibitem{r3} D.M. Herlac, A.B. Kaiser and J. Ustner, 1987 {\it Europhys. Lett.} {\bf  4} 97
\bibitem{r4} R. Harris and J. O. Strom-Olsen, 1983 {\it Topics in Appl. Phys.} {\bf  53} 325
\bibitem{r5} R.W. Cochrane and J. O. Strom-Olsen, 1984 {\it Phys. Rev. B} {\bf  29} 1088
\bibitem{r6} D. Markowitz, 1977 {\it Phys. Rev. B} {\bf  15} 3617
\bibitem{r7} B.Y. Boucher, 1972 {\it J. Non-cryst. solids} {\bf  7} 277
\bibitem{r8} Asad Niazi and A. K. Rastogi, 2001 {\it J. Phys.: Con. Mat.} {\bf  13} 6787
\bibitem{r9} L. Piraux, K. Amine, V. Bayot, J.-P. Issi, A. Tressaud and H. Fujimoto, 1992 {\it Solid state comm.} {\bf  82} 371
\bibitem{r10} S. Banerjee and A.K. Raychaudhuri, 1992 {\it Solid state comm.} {\bf  83} 1047
\bibitem{r11} J. Hrebik and B.R. Coles, 1977 {\it Physica B+C} {\bf 86} 169
\bibitem{r12} E. Rozenberg, M. Auslender, I. Felner, and G. Gorodetsky , 2000 {\it J. Appl. Phys.} {\bf 88} 2578
\bibitem{r13} A Barman, M Ghosh, S Biswas, S K De and S Chatterjee, 1998 {\it J. Phys.: Con. Mat.} {\bf  10} 9799
\bibitem{r14} P Pratumpong, R Cochrane, M A Howson and H-G Busmann, 2000 {\it J. Phys.: Con. Mat.} {\bf  12} 1805
\bibitem{r15} R. W. Cochrane, R. Harris, J. O. Ström-Olson, and M. J. Zuckermann  , 1975 {\it Phys. Rev. Lett.} {\bf 35} 676
\bibitem{r16} B. L. Altshuler and A. G. Aronov, 1979 {\it Sov. Phys. JETP} {\bf 50} 968 
\bibitem{r17} J. E. Graebner, J. J. Rubin, R. J. Schutz, F.S.L. Hsu, and W. A. Reed, 1975 {\it AIP Conf. Proc.} {\bf 24}, 445
\bibitem{r18} R L Rusby, 1974 {\it J. Phys. F: Metal Phys} {\bf 4} 1265
\bibitem{r19} Rikio Settai, Keisuke Katayama, Hiroshi Muranaka, Tetsuya Takeuchi, Arumugam Thamizhavel, Ilya Sheikin and Yoshisicko Onuki 2010 {\it Physics and Chemistry of Solids} {\bf 71} 700
\bibitem{r20} J. Labbe 1967 {\it Phys. Rev.} {\bf 158} 647
\bibitem{r21} C. Y. Huang, 1985 {\it J. Mag. Mag. Mat.} {\bf 51} 1
\bibitem{r22} R. H. Heffner and M. R. Norman, 1995	{\it arXiv:cond-mat/9506043v2}

\end{thebibliography}
\end{document}